\documentclass[aps,prl,twocolumn,superscriptaddress]{revtex4-1}

\usepackage{graphicx}
\usepackage{ulem}
\usepackage{color}

\usepackage{xspace}

\usepackage{amsmath}
\usepackage{siunitx}
\usepackage{miller}

\usepackage[version=3]{mhchem}
\usepackage{natbib}
\usepackage[]{hyperref}
\usepackage{ulem}
\usepackage{color}

\newcommand{\cro}{Ca$_{2}$RuO$_{4}$\,}

\renewcommand{\vec}[1]{{\boldsymbol #1}}
\newcommand{\expj}[1]{$j$\,=\,\SI[per-mode=symbol-or-fraction]{#1}{\ampere\per\square\centi\metre}}
\newcommand{\expE}[1]{$E$\,=\,\SI[per-mode=symbol-or-fraction]{#1}{\volt\per\centi\metre}}
\newcommand{\K}[1]{\SI{#1}{\kelvin}}
\newcommand{\Acm}[1]{\SI[per-mode=symbol-or-fraction]{#1}{\ampere\per\square\centi\metre}}

\newcommand{\add}[1]{{#1}}

\begin{document}


\title{Evidence for current-induced phase coexistence in \cro and its influence on magnetic order}

\author{K. Jenni}
\affiliation{$I\hspace{-.1em}I$. Physikalisches Institut,
Universit\"at zu K\"oln, Z\"ulpicher Str. 77, D-50937 K\"oln,
Germany}

\author{F. Wirth}
\affiliation{$I\hspace{-.1em}I$. Physikalisches Institut,
Universit\"at zu K\"oln, Z\"ulpicher Str. 77, D-50937 K\"oln,
Germany}

\author{K. Dietrich}
\affiliation{$I\hspace{-.1em}I$. Physikalisches Institut,
Universit\"at zu K\"oln, Z\"ulpicher Str. 77, D-50937 K\"oln,
Germany}

\author{L. Berger}
\affiliation{$I\hspace{-.1em}I$. Physikalisches Institut,
Universit\"at zu K\"oln, Z\"ulpicher Str. 77, D-50937 K\"oln,
Germany}

\author{Y. Sidis}
\affiliation{Laboratoire L\'eon Brillouin, C.E.A./C.N.R.S., F-91191 Gif-sur-Yvette CEDEX, France}

\author{S. Kunkem\"oller}
\affiliation{$I\hspace{-.1em}I$. Physikalisches Institut,
Universit\"at zu K\"oln, Z\"ulpicher Str. 77, D-50937 K\"oln,
Germany}

\author{C.~P.  Grams}
\affiliation{$I\hspace{-.1em}I$. Physikalisches Institut,
Universit\"at zu K\"oln, Z\"ulpicher Str. 77, D-50937 K\"oln,
Germany}

\author{D.~I. Khomskii}
\affiliation{$I\hspace{-.1em}I$. Physikalisches Institut,
Universit\"at zu K\"oln, Z\"ulpicher Str. 77, D-50937 K\"oln,
Germany}

\author{J. Hemberger}
\affiliation{$I\hspace{-.1em}I$. Physikalisches Institut,
Universit\"at zu K\"oln, Z\"ulpicher Str. 77, D-50937 K\"oln,
Germany}

\author{M. Braden}\email[e-mail: ]{braden@ph2.uni-koeln.de}
\affiliation{$I\hspace{-.1em}I$. Physikalisches Institut,
Universit\"at zu K\"oln, Z\"ulpicher Str. 77, D-50937 K\"oln,
Germany}





\date{\today}

\begin{abstract}

Combining quasistatic and time-resolved transport measurements with X-ray and neutron diffraction experiments
we study the non-equilibrium states that arise in pure and in Ti substituted Ca$_2$RuO$_4$ under the application of current densities.
Time-resolved studies of the current-induced switching find a slow conductance relaxation that can be identified with heating and a fast one
that unambiguously proves an intrinsic mechanism.
The current-induced phase transition leads to complex diffraction patterns.
Separated Bragg reflections that can be associated with the metallic and insulating phases by their lattice parameters, indicate a real structure with phase coexistence that strongly varies with temperature and current strength. A third contribution with a $c$ lattice constant in between those of metallic and insulating phases appears upon cooling.
At low current densities, this additional phase appears below $\sim$100\,K and is accompanied by a suppression of the antiferromagnetic order that otherwise can coexist with current carrying states. A possible origin of the intermediate phase is discussed.

\end{abstract}


\maketitle


\section{Introduction}

The interaction of magnetic, orbital, and lattice degrees of freedom in strongly correlated electron materials
frequently leads to the competition of different phases.
As a result small external stimuli can trigger phase transitions and exotic quantum states.
Especially the electric control of non-thermal phase transitions is of wide interest in view of possible applications \cite{Chudnovskiy2002,Vaju2008}.
It is, however, very difficult to establish a true non-equilibrium character of a new phase and to fully exclude  Joule heating or microscopic phase separation, which were shown to be relevant in many $3d$ transition-metal oxides \cite{Qazilbash2007,McLeod2017}.
For electric-field induced phenomena it was shown that the motion of defects can be essential \cite{Jeong2013,Szot2007}.
Therefore, a thorough understanding of the transition mechanisms is essential for the interpretation of any non-equilibrium phase.

In this paper, we investigate the antiferromagnetic (AFM) Mott insulator \cro , which is a "bad metal" (with low conductance) at high temperatures in thermal equilibrium and which transforms into a Mott insulator upon cooling below \K{357} \cite{Nakatsuji1997,Nakatsuji1999,Alexander1999}.
The metal-insulator transition is accompanied by strong structural changes \cite{Braden1998,Friedt2001,Steffens2005}.
At the transition the $c$ lattice parameter shrinks and in-plane parameters elongate, coupled with a transition
from elongated RuO$_6$ octahedra at high temperature to flattened octahedra in the insulating state. This seems to be
the essential element to change the orbital occupation and to induce insulating behavior \cite{Fang2001,Zhang2017}. But upon cooling also the
tilting of the RuO$_6$ octahedra increases and the octahedron basal plane becomes elongated along the orthorhombic $b$ direction \cite{Braden1998,Friedt2001,Steffens2005}.
This latter distortion seems to pin the magnetic moment parallel to the $b$ axis \cite{Kunkemoller2015}. The
strong structural changes are not restricted to the metal-insulator transition at \K{357} but extend over a large temperature interval down
to about the onset of AFM order at $T_{\mathrm N}$=110\,K. Note, that \cro \ is a layered material resulting in rather low
three-dimensional AFM ordering temperatures.

Besides by heating, the metal-insulator transition can be induced by applying hydrostatic pressure above 0.5 GPa \cite{Nakamura2002,Steffens2005}, by substitution for example with Sr \cite{Nakatsuji2000} and by applying an electric field \cite{Nakamura2013}. The electric field needed to drive the phase transition is unusually small (\expE{40}) compared to the Mott energy gap; it is about two orders of magnitude below a typical break-through field of a Mott insulator \cite{Whitehead1953,Taguchi2000}.
Following this discovery \cite{Nakamura2013} several other studies have confirmed current-induced quasi-metallic states in \cro \cite{Sow2017,Zhao2019,Okazaki2013,Zhang2019}.
In particular, by using a non-contact infrared thermometer it was shown that the metallic phase can be stabilized by an external current at a sample temperature well below the metal-insulator transition \cite{Okazaki2013}.
The low-temperature observation of the quasi-metallic states suggests a truly electric mechanism and thus excludes Joule heating of the entire sample as the origin \cite{Sow2017,Zhao2019,Okazaki2013}. However, if the sample becomes inhomogeneous and splits in metallic and insulating parts the heating becomes inhomogeneous as well.
Local heating remains a problematic issue in almost all experimental studies.

It was reported that \cro \ exhibits strong diamagnetism and anomalous transport properties \cite{Sow2017} at moderate current densities of 1 to \Acm{2}, but the same group recently showed that the strong diamagnetism is an experimental artefact \cite{Mattoni2020}.
At low temperature the magnetoresistance turns negative and the
Hall coefficient exhibits a sign change\cite{Sow2017}. Strong diamagnetism contrasts with  the ferromagnetic or quasiferromagnetic instabilities reported for the metallic phases reached by applying pressure or by substitution \cite{Nakamura2002,Friedt2004,Steffens2011,Steffens2005}.

For moderate current densities, nano-imaging optical techniques report nano-stripe structured areas of phase coexistence with different optical reflectivity \cite{Zhang2019}. This documents that the real structure of current carrying \cro \ can be non-homogeneous.
Diffraction experiments at a high current density of 10 A/cm$^2$ proposed new structural phases in a current carrying sample \cite{Bertinshaw2019}, but the crystal structures strongly resemble those \cro \ exhibits at higher temperatures and pressure \cite{Braden1998,Friedt2001,Steffens2005}.
Other diffraction experiments on a crystal with 3\% Mn substitution report evidence for a modified orbital arrangement \cite{Zhao2020}.
The latter study, furthermore, does not find any evidence for strong diamagnetism.

Here we report transport and single-crystal diffraction studies using neutron and X-ray radiation. The diffraction experiments with a careful recording of the sample temperature
indicate complex phase coexistence of at least three different phases distinguished by different $c$ lattice parameters.
In particular at low current densities, in addition to the initial metallic and insulating phases
an additional component appears.

\section{Experimental}

Single crystals of \cro were grown by the floating-zone technique and characterized by resistivity and magnetization measurements \cite{Kunkemoller2015,Kunkemoller2017}. Due to the metal-insulator transition at 357\,K and the accompanied structural transition the crystals tend to shatter into mm sized pieces upon cooling to room temperature in the furnace. A small amount of 1\% Ti substitution broadens the transition, which can result in large single crystals of up to 1 cm$^3$ volume.
Additionally and more importantly for this study, Ti-substituted crystals permit to pass the metal-insulator transition several times without destroying the sample.
The magnetism is not strongly influenced by the Ti, since it is isovalent to Ru and nonmagnetic \cite{Kunkemoller2017}. In the experiments presented  here crystals containing 1\% of Ti and pure ones were used.
For the application of the DC current in the quasistatic conductance and in the diffraction experiments, the plate-shaped crystals were glued to a copper plate using conductive silver paste. The $ab$ planes are parallel to the plate. The copper plate as well as the sample were contacted with copper wires, resulting in a current direction parallel to the $c$ axis. \add{We choose this current direction, because it offers the 
best conditions for thermalizing the sample thereby reducing heating issues. In addition cracking of the sample in thin plates
parallel to current is suppressed} 

Time-resolved pulsed transport measurements were performed in a high bandwidth coaxial setup suitable for frequencies up to the GHz range employing a current source [\textsc{Keithley 2400}] together with a 200\,MHz oscilloscope [\textsc{Agilent U2702A}] and a fast pre-amplifier [\textsc{StanfordResearch 560}].  The time-resolved temperature measurements were performed using a fast, thin (250\,$\mu$m) foil-type thermocouple [T-type] with a response time in the millisecond range. The sample was prepared as small platelet with a thickness of 0.5\,mm and a cross section of 0.8\,mm$^2$ covered with silver electrodes.

\begin{figure}
 \includegraphics[width=\columnwidth]{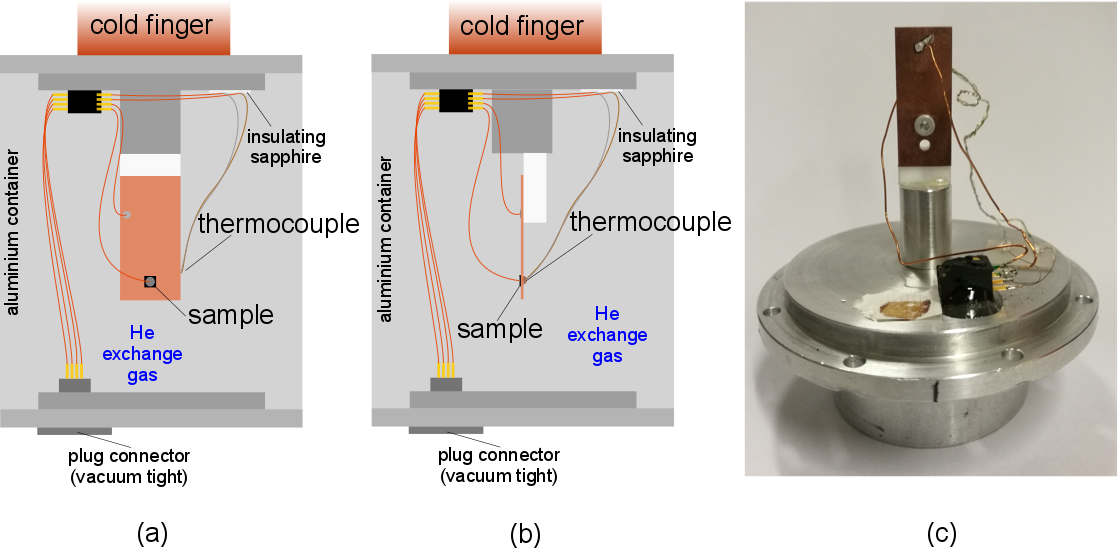}
  \caption{Illustration of the experimental setup to perform X-ray and neutron diffraction experiments on \cro \ with a current flowing
  parallel to the crystallographic $c$ direction (perpendicular to the Ru layers). Note, that the thermal couple opposite to the sample crystal
  allows for improved estimate of the sample temperature.
  }
  \label{mounting}
 \end{figure}

The X-ray diffraction (XRD) experiments were conducted on a D5000 powder diffractometer using Cu K$_\alpha$ radiation equipped with a Helium cryostat where the sample chamber is cooled by He flow.
Thermal contact to the single crystalline samples is guaranteed by the exchange gas inside the chamber.
For the neutron diffraction experiments the sample setup was incorporated into a sealed aluminum can including He exchange gas and a vacuum tight connector.
The can was then cooled using closed-cycle refrigerators. The elastic neutron scattering experiments were performed at the thermal two-axis diffractometer 3T1 at the Laboratoire L\'eon Brillouin (LLB).
Neutrons with a wavelength of 2.4\,\AA \ were extracted by a vertically focusing pyrolithic graphite (002) monochromator.
In order to achieve the high resolution required to distinguish the different contributions in \cro \ we used collimations of 15' and 10' before monochromator and detector, respectively, and higher order contaminations were suppressed with a pyrolithic graphite filter.
To apply the current and to read the voltage generated by the thermocouple a [\textsc{Keithley 2400}] source meter and a [\textsc{Keithley 2128}] nanovoltmeter were used.

To monitor the sample temperature as exactly as possible in the neutron experiment one contact of a thermal couple was placed on the backside of the 0.5\,mm thick copper plate, precisely at the sample position. As the reference temperature of the thermocouple the cryostat sensor was used, so that we can precisely determine the temperature at the sample position. The experimental mounting used in most of the neutron
experiments is shown in Fig.~1. In the X-ray diffraction experiment the thermocouple had to be mounted at the same side of the copper plate at a distance of 2\,mm from the crystal.


\section{Results}

\begin{figure}
\includegraphics[width=\columnwidth]{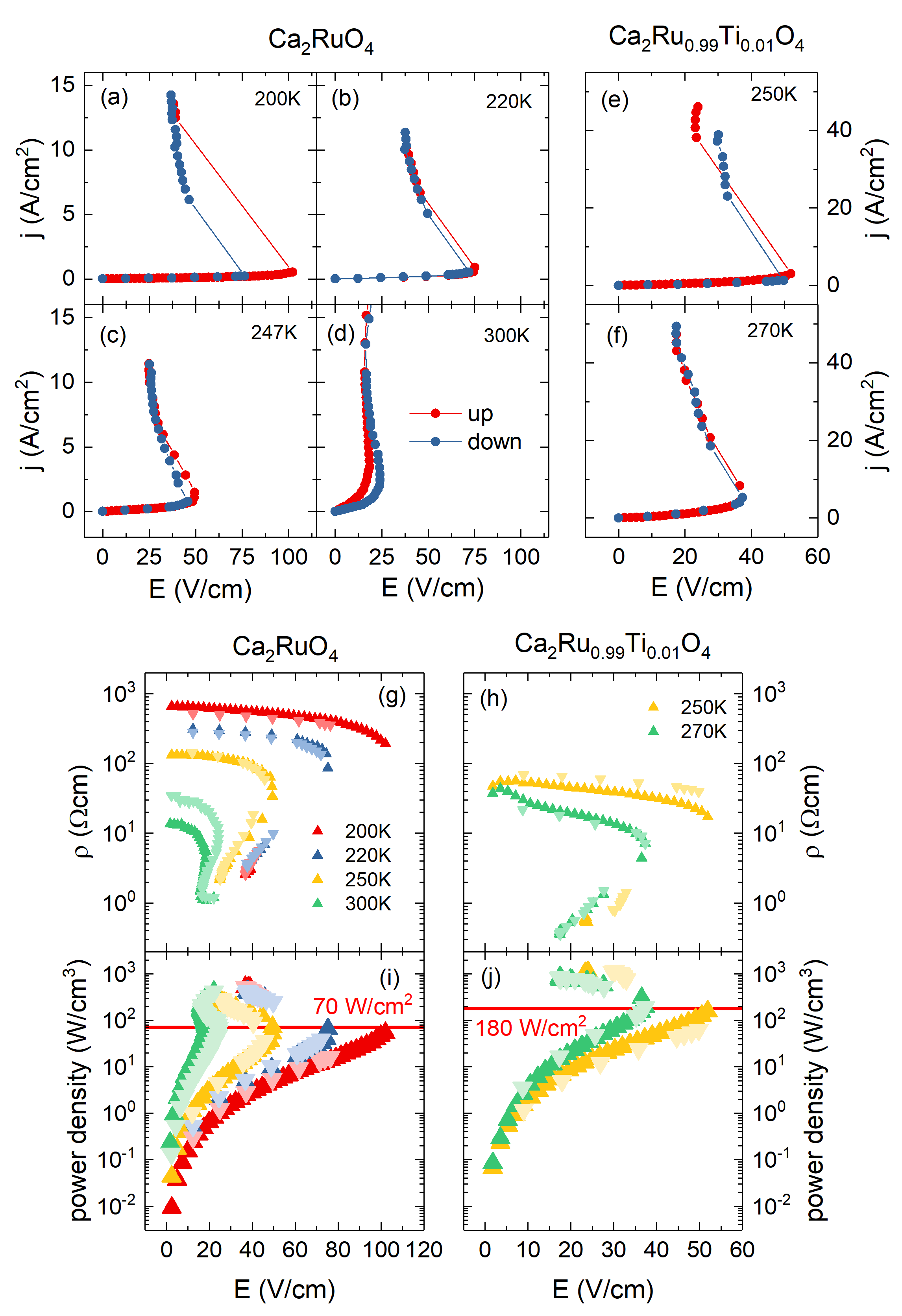}
\caption{Switching characteristics of pure and 1\% Ti-substituted \cro in dependency of temperature. (a)-(f) Characteristic $j$-$E$ curves display the current-induced phase transition in a voltage controlled setup (The electric field has been corrected for the circuit resistance effects). Here red coloring denotes data taken for ramping the voltage up and blue for ramping down. (g)-(h) The calculated resistivities from the two-point measurements above show a clear drop at a temperature dependent critical electric field. \add{(i) and (j) represent the corresponding power densities in the $j$-$E$ characteristic. The threshold power density, at which the switching occurs, is constant for all measured temperatures.} Dark coloring is used for the up direction, light coloring for the down direction.
}
  \label{switching}
 \end{figure}

\subsection{Quasi-static transport studies of the insulator metal transition}

Nakamura et al. \cite{Nakamura2013} found that the insulator-metal transition in \cro can be induced when applying an electric field of \expE{40} at room temperature. At this field the sample resistance drops leading to a jump in current density $j$.
For the transport studies we use a two-contact circuit with a finite pre-resistance that limits the current when the samples becomes metallic, and we drive the sample with controlled voltage.
In contrast, for the temperature-dependent diffraction studies we drive the crystal by controlling the current, as otherwise the current would strongly change with temperature.
When ramping the current up in the current-controlled mode the electronics regulates the voltage to reach the demanded current and initially exceeds the critical voltage or electric field and thus enforces the partial insulator-metal transition.
We confirm the current-induced insulator-metal transition \cite{Nakamura2013} in both samples, pure and Ti substituted ones, with critical electric fields of the same order (Fig.~\ref{switching}).
The critical field denotes the value, at which the material changes from insulating to conducting characteristics.
Upon cooling the sample resistivity increases and the critical electric field increases as well [Fig.~\ref{switching}(g),(h)] similar to reports in \cite{Nakamura2013,Cirillo2019}.
The more insulating the sample becomes, the higher fields are necessary to transform it to the metallic state.
For this reason any changes of current or voltage in the XRD and neutron experiments were applied above \K{250}, where the samples are still conductive enough and where stronger cooling power is provided.

While the critical electric field increases, the critical current density decreases with decreasing temperature.
The resulting \add{critical} power density does not exhibit a temperature dependence but stays constant \add{for the measured temperature range between 220 and 300 K [Fig.~\ref{switching}(i), (j)].}
which indicates that the current-induced insulator-metal transition is not simply due to heating of the entire crystal.
Many aspects of the electric-field-driven insulator-metal transition in \cro \ resemble that in VO$_2$, for which it was also
reported that the threshold power density does not vary with temperature \cite{Mansingh1980}.
Both \cro \ samples, pure and substituted ones, show comparable switching behavior. The substituted sample exhibits a slightly lower resistivity which is connected to higher critical current densities. The  \add{threshold} electrical power density is therefore more than a factor two higher than that in the pure sample. [Fig.~\ref{switching}(i), (j)].
Nevertheless the Ti substitution does not effect the current-induced insulator-metal transition significantly, for which reason the substitution can be disregarded for the following analysis of results.

The form of the $j$-$E$ characteristic, see Fig.~\ref{switching}, that has consistently been reported by many groups \cite{Nakamura2013,Okazaki2013,Cirillo2019} inevitably leads to the formation of inhomogeneous current-carrying states. This effect has been shown on general grounds, and it is well known in the physics of semiconductors \cite{Volkov1969} that the state with negative differential resistance $\frac{dI}{dV}<0$  is absolutely unstable. Therefore, the $S$-shaped $j$-$E$ (or $I$-$V$) characteristic leads to the formation of filaments predominantly parallel to the current (but in real system usually forming a percolation network),  whereas the N-shaped $I$-$V$ curve results in the formation of insulating and metallic domains (Gunn domains) perpendicular to the current  (Gunn domains usually move with current leading to the oscillating behavior - Gunn oscillations). The formation of metallic filaments in an insulating matrix is well documented in many systems with insulator-metal transitions, notably in VO$_2$ \cite{Kumar2013,Mansingh1980,Berglund1969,Zhang2014} and in SrTiO$_3$\cite{Stille2012}; they lead to a percolation picture of conduction and to switching phenomena. Due to the impact of strain, however, the arrangement of the metallic parts can essentially change, see discussion below.

As a result of the inhomogeneous state the current is predominantly concentrated in narrow metallic channels, in which the local current density is much higher than the average one, so that the local Joule heating $jEV = \rho j^2$ (in the current-controlled regime, as used in most experiments) and the resulting local  temperature in the current carrying parts will also be higher than that estimated by bulk probes. 
The question whether the metal-insulator switching often observed in systems with metal-insulator transition is due to intrinsic effects or caused by local heating is difficult to answer.

\subsection{Time-resolved analysis of the current-induced insulator-metal transition}

\begin{figure}
\includegraphics[width=0.8\columnwidth]{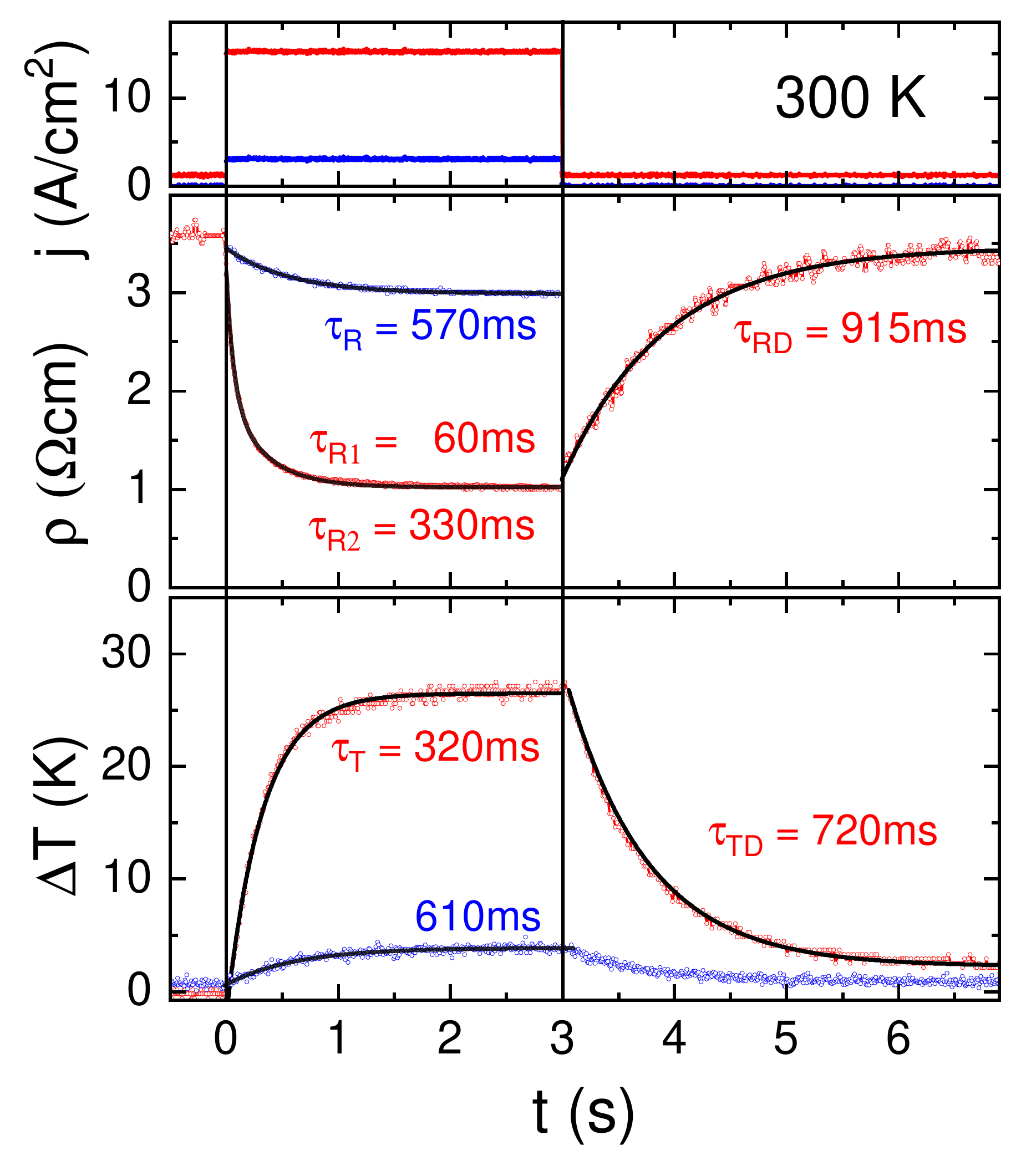}
\caption{Pulsed current measurements of sample resistivity (middle panel) and temperature (bottom) in Ca$_2$RuO$_4$.  For driving currents below the threshold (3\,A/cm$^2$, blue, top panel) the resulting change in  $\rho$ can be modeled with an mono-exponential decay with a decay time similar to the rise time found for the concomitant heating (bottom panel). For larger driving currents above the threshold (15\,A/cm$^2$, red, top panel) the change in $\rho$ only can be modeled employing a bi-exponential decay: A slowly relaxing component $\tau_{R2}\approx 300$\,ms in accord with the rise time of temperature (bottom) and a significantly faster process with $\tau_{R1}\approx 60$\,ms, which seems to be directly induced via the current or, correspondingly, via the electric field.
}
  \label{fig_pulse}
 \end{figure}

In order to elucidate the question what induces the rise in conductivity, the electric field connected to the forced current or simple Joule heating, we performed time-resolved pulsed experiments as illustrated in Fig.~\ref{fig_pulse}. Most likely one has to consider phase separation into an insulating and a percolating phase carrying most of the current. The idea of the time-resolved study is to disentangle time scales of a field-driven switching process and of presumably unavoidable heating effects.

The upper panel of Fig.~\ref{fig_pulse} displays the pulse pattern used to induce additional conductivity. Switching-on a smaller current of about 3\,A/cm$^2$ leads to a mono-exponential drop in resistivity (middle panel) of about 15\%\ with a slow time constant of $\tau_{R}=570$\,ms. At the same time the sample temperature rises with a similar time constant saturating 3.2\,K above the starting value of 298\,K. Obviously, the resistivity changes in this below-threshold regime are coupled to the heating. After switching-off the current at $t = 3$\,s the temperature decays to 298\,K again. Considering a power dissipation of about $\rho j^2 \approx$15\,W/cm$^3$, and thus still of the order of 6\,mW in the sample crystal even for this smaller driving current, such heating has to be expected. The red curves in Fig.~\ref{fig_pulse} denote results for the larger driving current of 15\,A/cm$^2$. Here a small base current of 1\,A/cm$^2$ is applied before and after the pulse to enable the determination of the resistivity in these regimes. The drop in resistivity now amounts up to 70\%\ but it is not possible to describe the underlying time dependence using only one relaxation time. A bi-exponential fit reveals a much faster decay time of $\tau_{R1}\approx 60$\,ms, together with a slower one of $\tau_{R2}\approx 330$\,ms. The latter is in accord with the corresponding rise time of the temperature (bottom). The faster process seems not to be induced by simple heating (even though a considerable increase in temperature of up to 28\,K can be monitored). This second mechanism has to be attributed to the direct induction of a more conductive phase via the current or, correspondingly, via the electric field. It is interesting to note that after switching-off this larger driving current, the electrical relaxation time exceeds the thermal one. The field-induced phase appears to be metastable which is in accord with the observed hysteresis in the $j$-$E$ curves, see \cite{Nakamura2013} and Fig.~2.

The observation of two relaxation rates in the resistance unambiguously confirms an intrinsic origin of the current induced switching.
There are several processes that can be associated with the faster conductance enhancement, such as a purely electronic mechanism or
filament formation. The fact that even this faster process happens on a time scale above milliseconds suggest some structural implication.

\subsection{X-ray and neutron diffraction studies as function of current density and temperature}

Since the metallic and insulating phases in \cro differ strongly in their lattice constants, diffraction is a suitable experimental technique to investigate phase changes in this material. Especially the lattice parameter $c$ strongly increases by $\approx 3\%$ between the insulating phase at room temperature and the metallic phase at $\sim$\K{360} \cite{Braden1998,Friedt2001}, which allows one to distinguish these two phases. However, a good resolution is required in neutron diffraction experiments.
One expects to observe (00L) reflections at lower $2\theta$ values when the material becomes metallic. The metallic phase (MP) of \cro \ was initially labeled as $L$ phase due to its longer $c$ axis, while the insulating phase (IP) is frequently labeled as $S$ (for short) phase \cite{Nakatsuji1997,Braden1998}.

As described above, switching the \cro crystals from the insulating to the metallic state involves large power densities. Note that a power density of 100\,W/cm$^3$ would
result in a temperature drift of 36\,K/s for a material without any thermal contact just by taking the specific heat into account \cite{note-cp}.
As illustrated in the time-resolved experiments, heating of the sample during the diffraction studies cannot be completely avoided and must be carefully taken into consideration. The controversial diffraction results \cite{Nakamura2013,Bertinshaw2019,Zhao2020,Cirillo2019} most likely stem from differing heating conditions. With an additional sensor placed close to the crystal and with the good electric and thermal contact of the sample to a large Cu plate heating effects could be reduced and better documented in most of our experiments.

\begin{figure}[htbp]
	\begin{center}
	\includegraphics[width=\columnwidth]{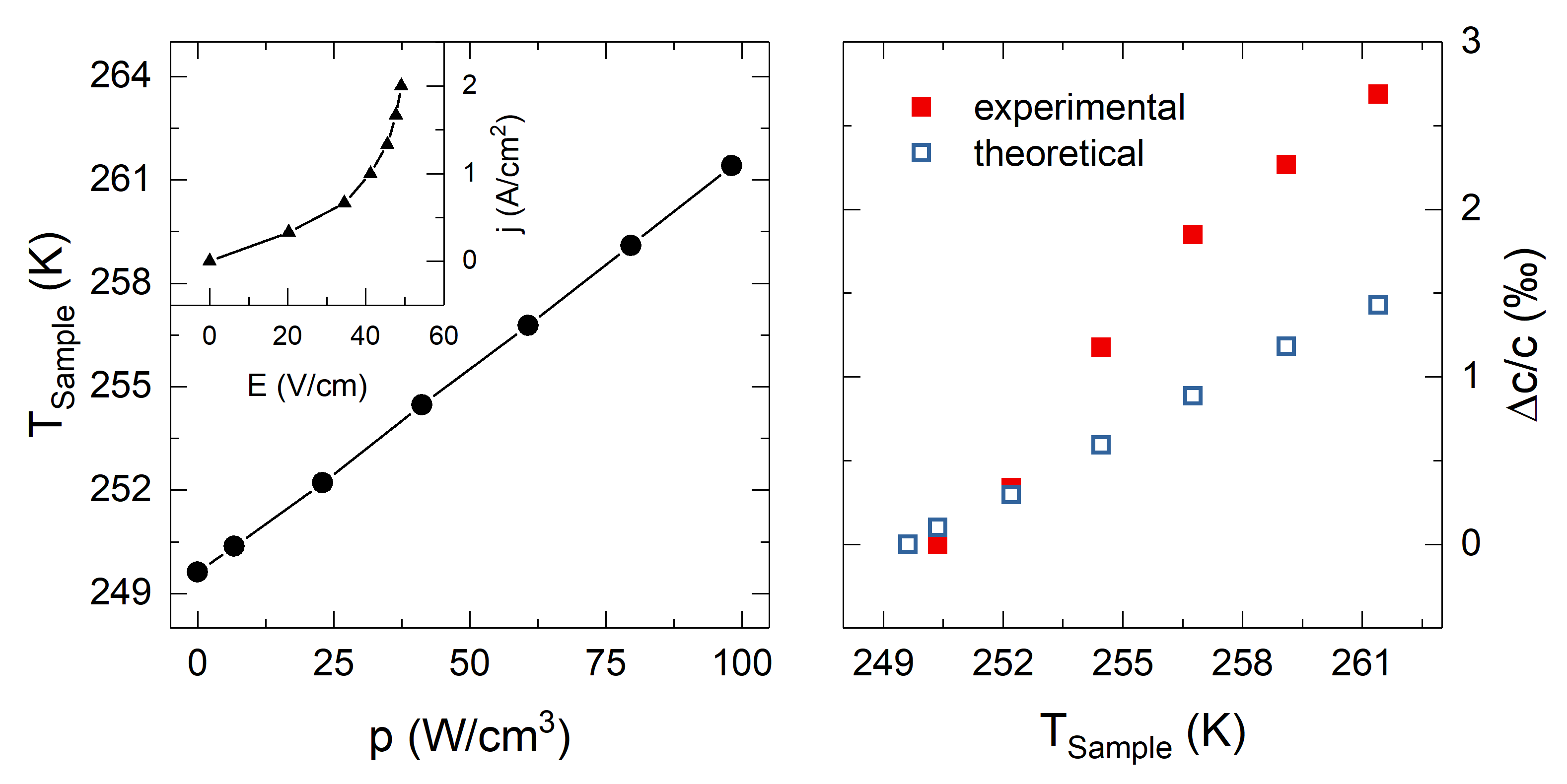}
	\caption{Sample temperature (thermocouple) in comparison to power density and relative lattice change while ramping up current at the fixed cryostat temperature of \K{250}. The purely temperature-driven relative lattice change (open squares) is taken from \cite{Friedt2001} for the given temperature range. The experimental lattice change is derived from the (006) reflection in the neutron diffraction experiment.}
	\label{ramping}
	\end{center}
\end{figure}

As a first example, we show the ramping up of the current density in a neutron diffraction experiment in Fig.~\ref{ramping}. The cryostat temperature was set to \K{250}, which is advantageous compared to simple experiments at ambient conditions, because it provides strong cooling power. The thermocouple with its sensor placed opposite to the crystal on the Cu plate shows
that heating occurs even at these rather low current densities. Note that the thermocouple can only give a lower estimate of the true sample temperature. Similar to several other studies we observe an increase of the $c$ lattice parameter of the insulating phase. Half of this can be attributed to the temperature change detected in the thermal couple and thus to heating of the entire sample, but most likely an even larger part of the $c$ parameter enhancement is simply due to heating. In the scenario of phase separation the metallic
parts will sense even higher local heating.

\begin{figure}[!ht]
 \includegraphics[width=\columnwidth]{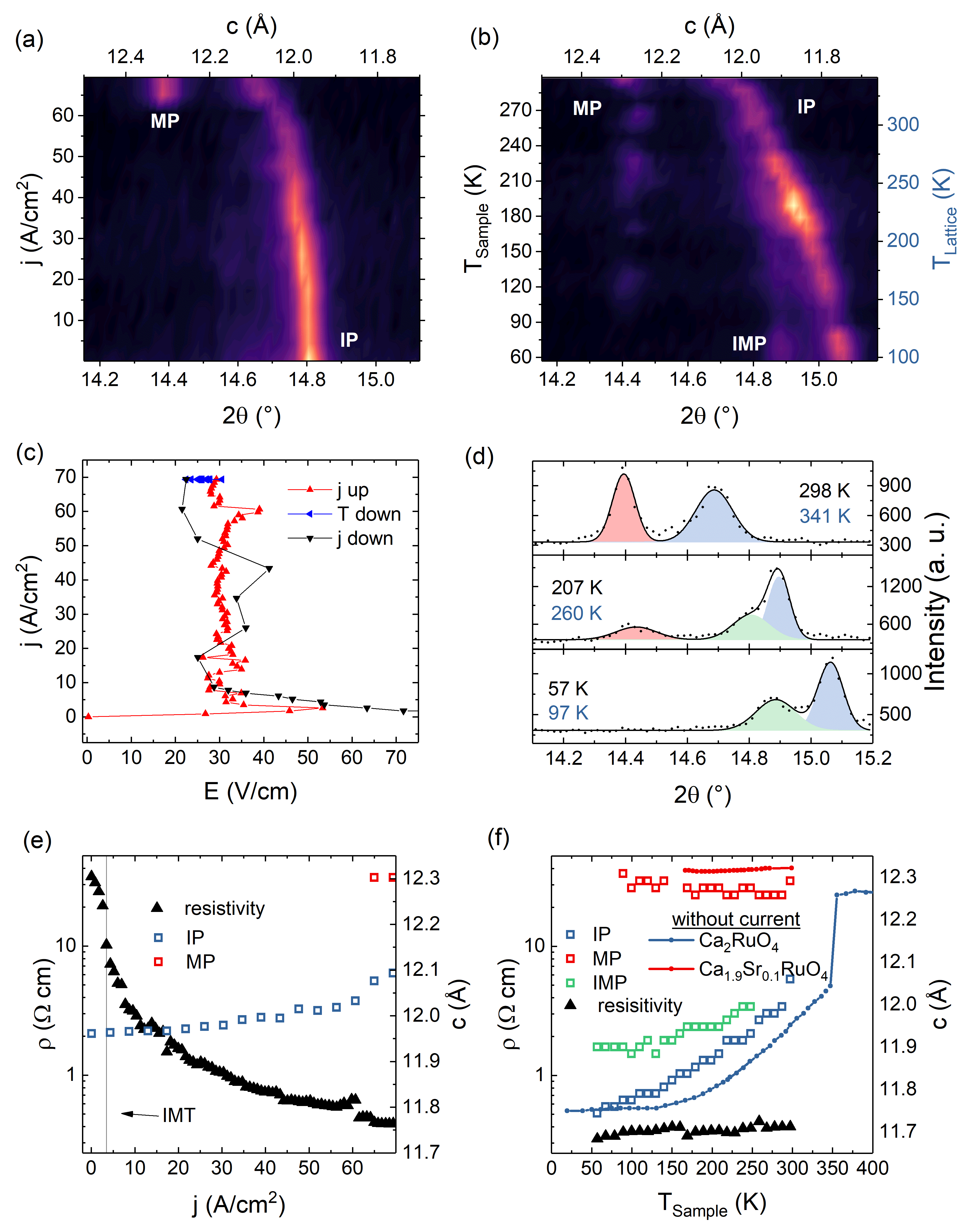}
	  \caption{X-ray diffraction study of the (002) reflection under an applied current density of up to \expj{69} on pure \cro. Panels (a) and (b) show the evolution of different phases characterized by different $c$ lattice parameters while ramping up current and cooling down with constant current, respectively. The coexistence of phases, visible in the three reflections, is strongly temperature dependent. (c) The $j$-$E$ curve is measured simultaneously to the XRD experiment while ramping up current, cooling down and decreasing the current. (d) The quantitative analysis of $2\theta$ scans clearly displays the phase mixture of three phases. Here the metallic (MP) and insulating (IP) phase are colored in red and blue, respectively. The third phase is colored in green and is labeled as intermediate phase (IMP). $T_{\mathrm{Sample}}$ (black) represents the cryostat temperature corrected by the thermocouple while $T_{\mathrm{Lattice}}$ (blue) is calculated by comparing the reflection angle of insulating phase with the temperature dependency of a current free sample, taken from \cite{Friedt2001}, i.e. ignoring any intrinsic change. (e),(f) Resistivity and lattice parameter in dependence to current density and sample temperature, respectively. The resistivity displays the occurrence of the insulator-metal transition long before the metallic phase becomes visible in the XRD measurement (e) and the decreasing resistivity while cooling down coincides with the phase existence of the intermediate phase (f). The temperature dependency of $c$ lattice parameter of \cro, from \cite{Friedt2001}, is added for comparison. The low temperature lattice parameter of the metallic phase (red line) is taken from data of Ca$_{1.9}$Sr$_{0.1}$RuO$_{4}$\, from the same reference.
  }
  \label{xrd}
\end{figure}

Some studies seem to partially ignore the heating and report an important intrinsic change of the insulating phase with small currents, a modified short-$c$  phase. However, this modified short-$c$ phase in Ref. \cite{Bertinshaw2019} exhibits a $c$ lattice constant increase just up to the maximum value that the insulating \cro \ phase reaches upon heating without a current (before the insulator-metal transition takes place at \K{357}). Furthermore, the modified short-$c$ phase proposed at \K{130} with a large current density of \Acm{10} perfectly agrees with the current-free insulating phase at the temperature of 240\,K:
The lattice constants reported in \cite{Bertinshaw2019} amount to $a$=5.404, $b$=5.547, and $c$=11.848\,\AA \ compared to $a$=5.407, $b$=5.560, and $c$=11.854 \,\AA \ for current-free  \cro \ at 240\,K, taken from \cite{Friedt2001}, and also the internal parameters perfectly agree \cite{note-struc}.


XRD experiments with a Ti-free sample show a second (002) reflection appearing when the current density is increased to \expj{69} at a temperature of \K{294} \add{[Fig.~\ref{xrd} (a)]}.
Note, however, that the sample was already mounted into a cryostat \add{and temperature stabilized} in order to reduce the local heating.
The characteristic $j$-$E$ curve \add{[Fig.~\ref{xrd}(c)] and the resistivity [Fig.~\ref{xrd}(e)]} indicate that the sample becomes already conductive at much lower current densities ($\approx \Acm{2.6}$), where no significant change in the diffraction pattern is observable \add{[Fig.~\ref{xrd}(a)]}.
At high temperature, the Bragg reflection associated with the metallic phase exhibits a similar intensity as that of the insulating phase, but it becomes rapidly weaker upon cooling [Fig.~\ref{xrd}(b),(d)] in spite of the rather large current flowing.
The $c$ lattice parameter of the metallic phase stays constant upon cooling in agreement with measurements on Sr-substituted materials, in which the
metal-insulator transition occurs at lower temperatures \cite{Friedt2001} \add{Fig.~\ref{xrd}(f)]}.
The $2\theta$ value of the Bragg peak associated with the insulating phase increases upon cooling following the pronounced shortening of the $c$ lattice parameter of the insulating phase at low temperature.
The cryostat temperature has been corrected by the offset indicated by the thermal couple scaled by a factor 2.5 in order to take the slight displacement of the thermocouple in the X-ray diffraction experiment into account.
Nevertheless the observed $c$ values of the insulating phase lie above the reported temperature dependence of the insulating phase without any current \add{[Fig.~\ref{xrd}(f)]}.
At high temperature, the insulating phase reflection exhibits significant broadening, see Fig.~\ref{xrd}(d), which can be easily recognized from the comparison with the metallic phase peak \add{[Fig.~\ref{xrd}(d)]}. Below $\sim$\K{260} a third reflection appears that unambiguously indicates phase coexistence of at least three phases with different $c$ lattice parameters: besides the insulating phase (IP) and the metallic phase (MP) there is an intermediate phase (IMP), see Fig.~\ref{xrd}(d).

\begin{figure}[hb]
\includegraphics[width=\columnwidth]{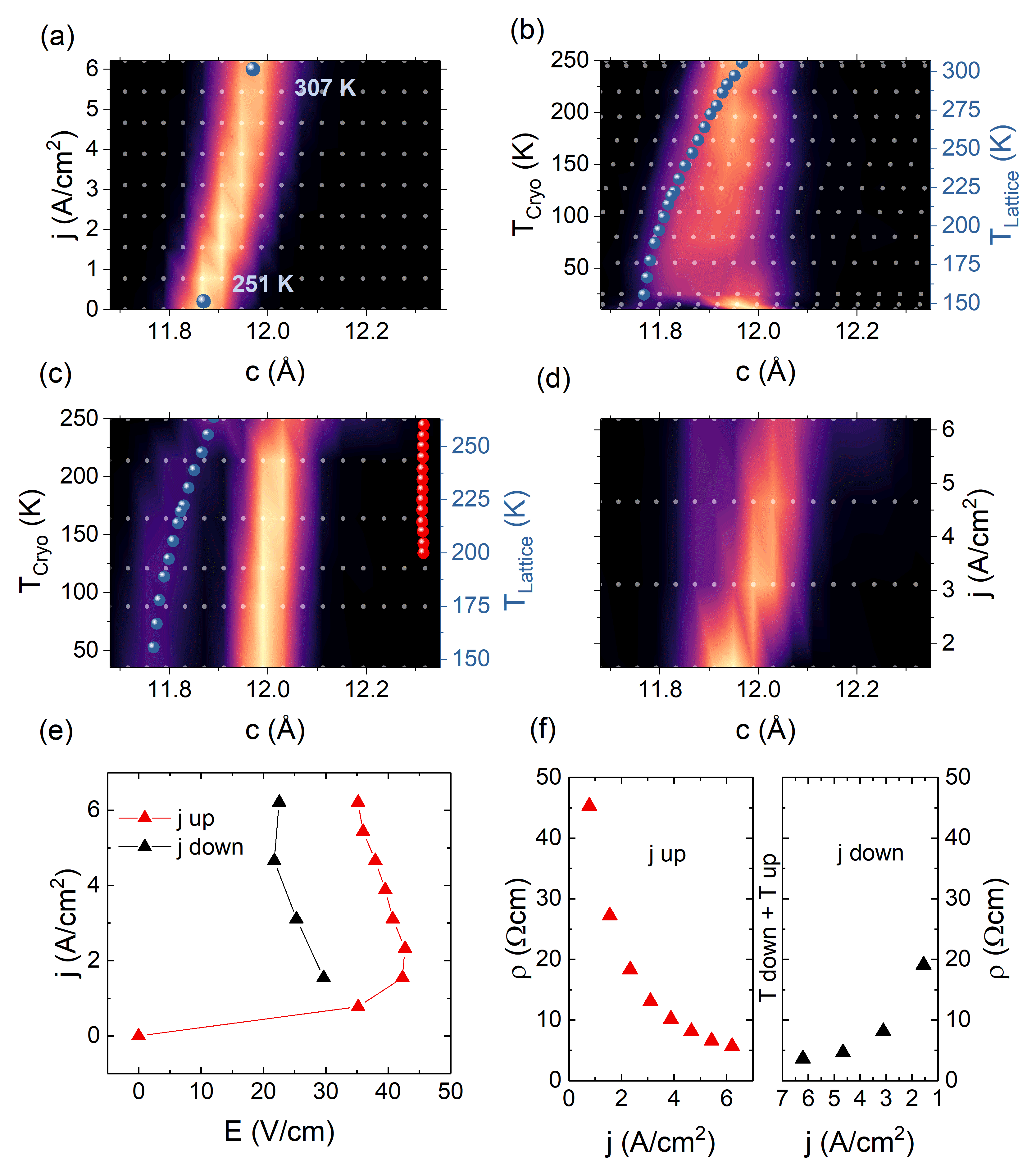}
  \caption{Evolution of $c$ lattice parameter recorded by elastic neutron scattering in a measuring sequence of increasing current (a), decreasing temperature (b), increasing temperature (c), and decreasing current (d). Longitudinal scans are taken across the (006) reflection in pure \cro at various current densities and temperatures. The blue circles represent the temperature dependency of \cro \ lattice parameters without current, taken from \cite{Friedt2001} and is used to determine $T_{\mathrm{Lattice}}$ from the experimental lattice parameter of the insulating phase. The low temperature lattice parameter of the metallic phase (red balls) is taken from data of Ca$_{1.9}$Sr$_{0.1}$RuO$_{4}$\, from the same reference and plotted against $T_{\mathrm{Lattice}}$. The small grey dots represent the measured points \add{and the color maps are obtained with an interpolation algorithm}.(e) The $j$-$E$ curve is measured simultaneously to the neutron diffraction experiment while ramping up current, cooling down, heating up and decreasing the current. (f) Sample resistivity is derived from the two-point measurement and displayed for the measurement sequence.
  }
  \label{tjmaps}
 \end{figure}

\begin{figure*}[htbp]
 \includegraphics[width=\textwidth]{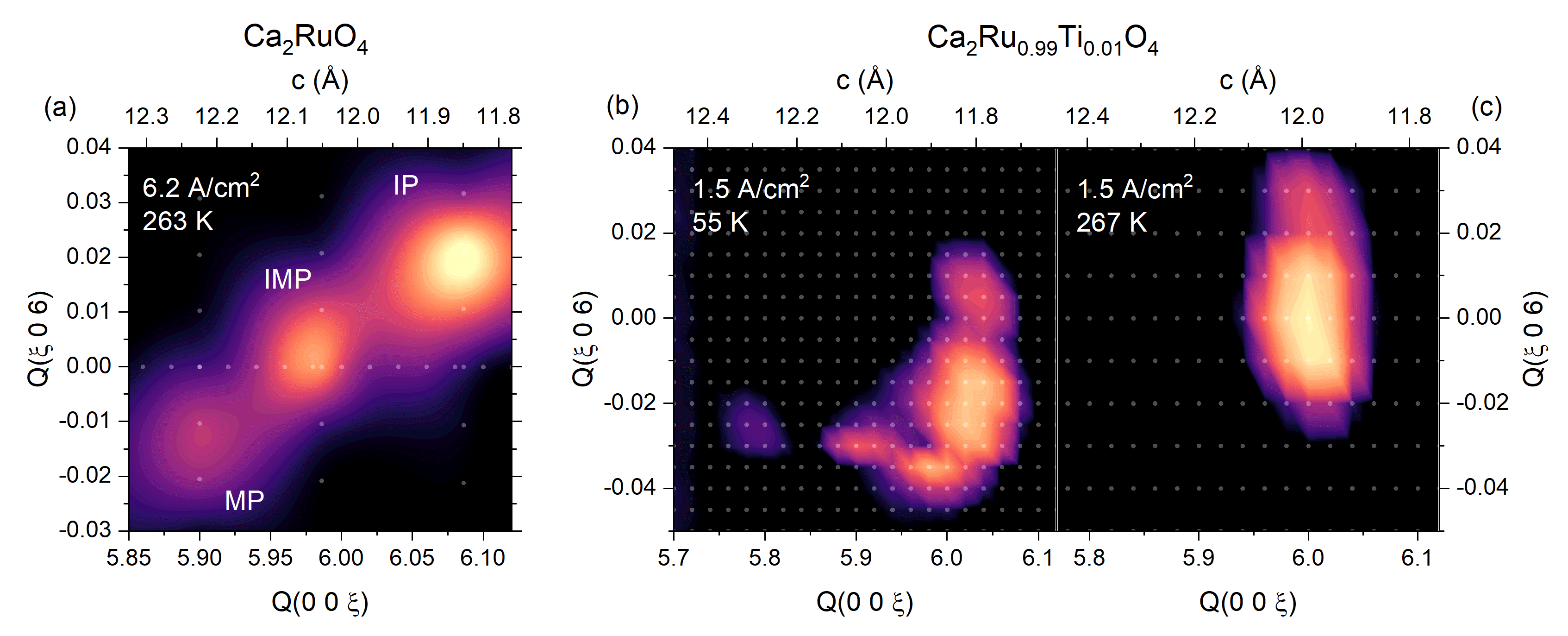}
  \caption{Mapping of reflections around (006) Bragg peak for two different samples, temperatures, and current densities. The small grey dots represent the measured points \add{and the color maps are obtained with an interpolation algorithm}.
  }
  \label{qmaps}
\end{figure*}

The coexistence of three phases is confirmed by neutron diffraction experiments that also aimed to characterize the magnetic ordering in the low current-density range.
In a first set of experiments \add{(Fig.~\ref{tjmaps})} the current density was ramped up to \expj{6.2}. An additional thermocouple could not be used in this experiment.
The color maps in Fig.~\ref{tjmaps}(a)-(d) represent the lattice parameter $c$ extracted from longitudinal scans across the (006) Bragg position and its change during a full measurement cycle of ramping up the current, cooling, reheating, and ramping the current down.
At a cryostat temperature of \K{250} the current increase leads to an increased $c$ lattice parameter while no additional reflection becomes visible \add{[Fig.~\ref{tjmaps}(a)]}. The $j$-$E$ curve clearly indicates a transition from insulating to metallic state at only \expj{1.55} , see Fig.~\ref{tjmaps}(f).
The shift of the insulating phase reflection can mostly be assigned to a thermal drift of the sample and therefore it can be used to roughly estimate the sample temperature, although this will overestimate the temperature due to the neglecting of the intrinsic shift, see discussion below.
A current density of \expj{6.2} heats up the sample by $\sim \K{50}$.
When cooling the sample under fixed current a phase separation at low temperatures occurs similar to the XRD observation.
The significant broadening of the insulating phase peak increases until two reflections can be distinguished, which we assign to the insulating phase and the intermediate phase \add{[Fig.~\ref{tjmaps}(b)]}.
Interestingly this phase segregation is not reversible as the two distinct reflections stay separated when increasing the temperature again, see Fig.~\ref{tjmaps}(c).
The phase separation and the appearance of the intermediate phase is associated with a reduction of the resistance, see Fig.~\ref{tjmaps}(e) and (f), which also
reduces the sample heating. Therefore, the sample temperature with the cryostat stabilized at \K{250} is lower after the cooling and heating cycle.
In addition to the insulating and intermediate phase reflections a weak third peak becomes visible at high temperatures with a $c$ lattice parameter close to the metallic phase.
This third phase also disappears quickly when reducing the current density at $T_{\mathrm{cryo}}$ = 250\,K at the end of the full cycle. Below \Acm{3} the intermediate phase reflection fades out and only one reflection persists [Fig.~\ref{tjmaps}(d)].
The presented full measurement cycle documents a complex real structure arising from phase segregation with strong hysteresis.
The amount of insulating, intermediate and metallic phase are not only determined by current strength and temperature but also depend on the history of the sample, similar to reports in \cite{Zhao2020}.

The mappings of the reflection against longitudinal (00$\xi$) and transversal ($\xi$00) directions clearly reveal the phase coexistence (see Fig.~\ref{qmaps}).
At high current densities the phase separation is already visible at high temperatures whereas at low current densities the coexistence only appears at low temperatures.
It can be concluded that the samples show more than the two conventional insulating and metallic phases at low temperatures.
The different positions of the Bragg peaks in the transverse direction (see Fig.~\ref{qmaps}) indicate that the different lattices are slightly tilted against each other similar to ferroelastic domains in a martensitic transition, see e.g. \cite{Khachaturyan1983,Khachaturyan1991}.

\begin{figure}
 \includegraphics[width=\columnwidth]{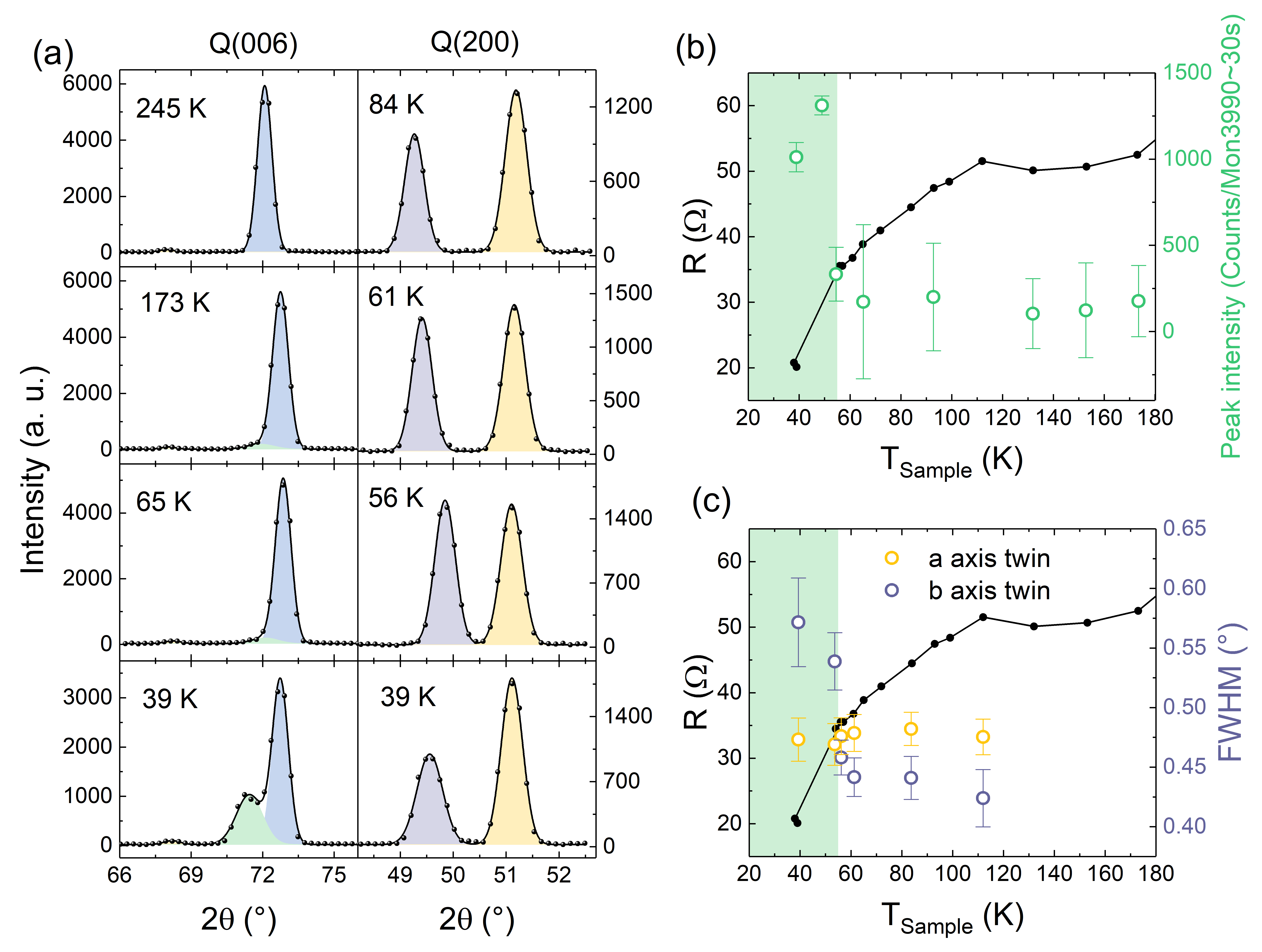}
  \caption{(a) Longitudinal scans across the (006) Bragg peak, left column, and (200) Bragg peaks, right column, at various temperatures and with applied current (\expj{1.5}) on pure \cro. The black line represents the sum of Gaussians while the single contributions are marked by the colored areas. In the left column the blue area illustrates the metallic phase, green the intermediate phase whereas the different twin fractions visible in the longitudinal scans of (200) are displayed in yellow ($a$ axis twin) and purple ($b$ axis twin). (b) Comparison of the temperature dependence of the sample resistance and the fitted peak intensity of intermediate phase. (c) Comparison of the temperature dependence of the resistance and the fitted peak width of both twin fractions.
  }
  \label{neutrons}
 \end{figure}

Using a two-axis diffractometer the accessible $\vec{Q}$ space is limited to a scattering plane spanned by two crystallographic directions. Therefore, a full structural analysis is not possible. But the twinning of the crystals with respect to the orthorhombic distortion allows one to measure the (200) and (020) reflections in addition to the (006) reflection [Fig.~\ref{neutrons}(a)]. In this set of neutron experiments with a stronger closed cycle refrigerator [{\sc Sumitomo RDK 205D}] and with a thermocouple indicated in Fig.~\ref{mounting} we cooled the crystal with a further reduced current density of \expj{1.5}. We find qualitatively the same phase segregation and the appearance of the intermediate phase reflection below $\sim$\K{50}, as it is shown in the longitudinal scan across the (006) Bragg position, see Fig.~\ref{neutrons}(a).
The longitudinal scans across (200) show two reflections. The $a$-axis twin results in higher $2\theta$ values and the $b$-axis twin in lower values, respectively, because the lattice parameter $a$ is smaller than $b$. The $a$-axis peak does not exhibit a strong temperature dependence, neither in $2\theta$ nor in peak width. However, the $b$-axis reflection shifts to higher $2\theta$ values synonymous with the decrease of the lattice parameter $b$. Only at \K{39} the lattice parameter $b$ increases again. Additionally to the shift, the Bragg peak of the $b$-axis twin is significantly broadened  below \K{50} \add{[Fig.~\ref{neutrons}(c)]}. This broadening can be attributed to the appearance of the intermediate phase, whose $b$ lattice parameter is expected to be slightly smaller since it exhibits a more metallic character. Both, the appearance of the intermediate phase and the broadening of the $b$-axis reflection below \K{50}, coincide with a drop in sample resistance [Fig.~\ref{neutrons}(b),(c)]. This again connects the appearance of the intermediate phase to an increased conductivity.

\begin{figure}
 \includegraphics[width=\columnwidth]{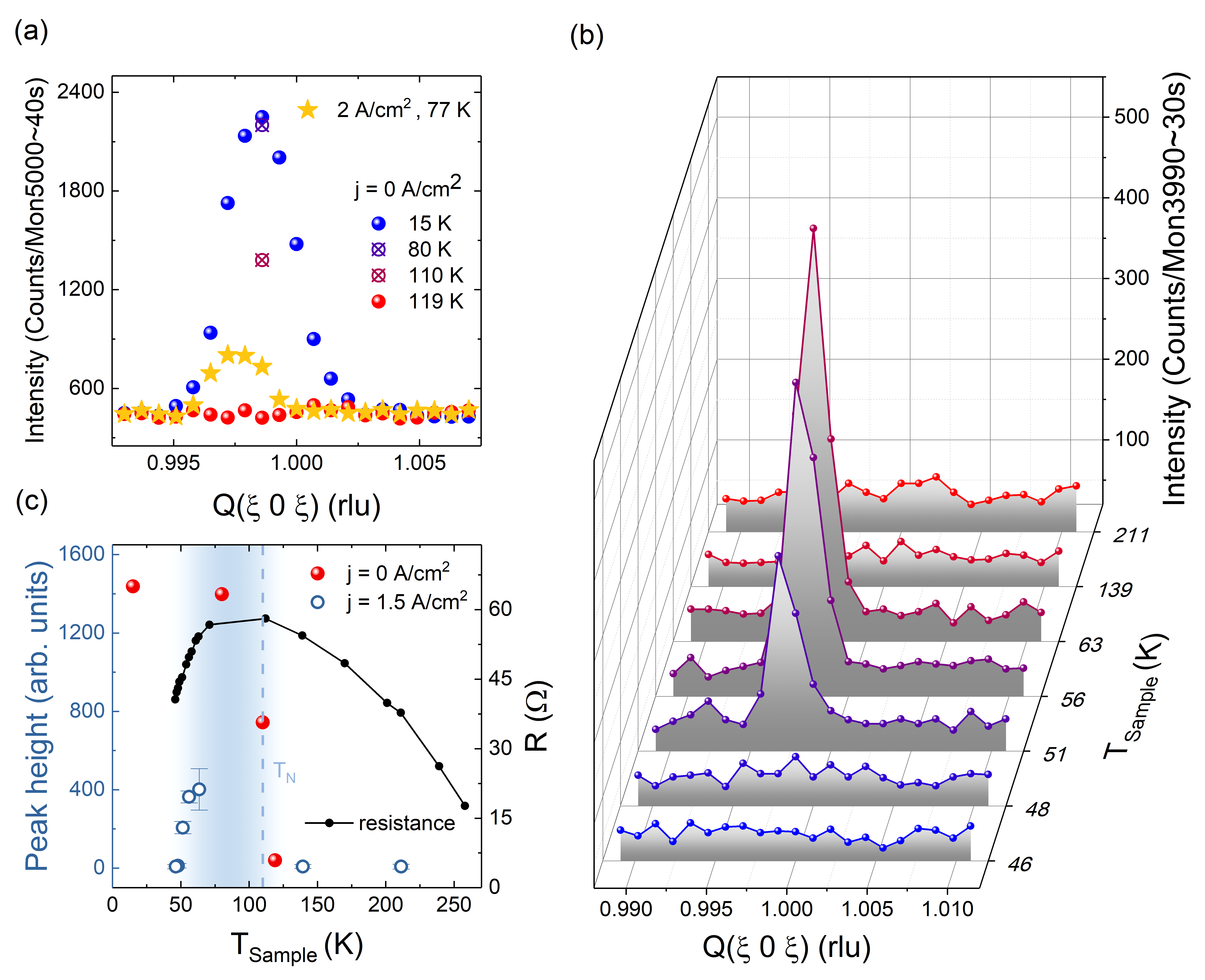}
  \caption{Influence of current application on the magnetic (101) reflection in substituted \cro. (a) Longitudinal scans over (101) and single-point count rates at different temperatures. The magnetic signal is reduced but persists with a low current density of \expj{1.5}. (b) Temperature dependence of longitudinal scans over the (101) reflection with an applied current density of \expj{1.5}. (c) Temperature dependence of the sample resistance recorded simultaneously with the neutron scattering in comparison with the peak heights from panel (b) (obtained by fitting with Gaussians). The red spheres represent the peak height of the (101) reflection measured without current from panel (a).
  }
  \label{magnetism}
 \end{figure}

Sow et al. showed that below \K{50} \cro exhibits anomalous properties under applied current  \cite{Sow2017}. In their magnetization measurements they could not find any antiferromagnetic transition for nonzero current densities.  Ti substituted \cro exhibits the B-centered AFM order \cite{Kunkemoller2017}, which leads to a magnetic (101) Bragg reflection \cite{Kunkemoller2017}. We studied this reflection depending on temperature and under applying a small current density (Fig.~\ref{magnetism}).
Without current application the peak intensity rapidly decreases only close to the transition temperature T$_N\,=\,110$\,K and becomes zero already at \K{119} [Fig.~\ref{magnetism}(a)]. Single point counts on the maximum of the longitudinal scan over (101) at \K{80} show that the intensity has little changed up to this temperature in agreement with previous reports \cite{Braden1998,Kunkemoller2017}. While applying a current of 120\,mA (\expj{2}) the lowest sample temperature reached is \K{70} as determined by the thermocouple. At this experimental configuration ($T_{\mathrm{sample}}$\,=\,\K{77}, \expj{2}) the (101) reflection remains visible, although with reduced intensity compared to the current-free phase at \K{80}. It can be clearly stated that magnetically ordered regions survive the application of small currents. The reduction of peak intensity can be caused by various factors such as reduction of ordered phase volume or reduction of the transition temperature and ordered moment.
By applying lower current densities (\expj{1.5}) it is possible to cool the sample below \K{50} (as indicated by the thermocouple). Figure \ref{magnetism}(b) shows the temperature dependent evolution of the (101) reflection. The peak intensities in (c) reflect the appearance of the magnetic Bragg peak at \K{63}, well below the transition temperature of current free \cro .
Surprisingly the Bragg peak at the magnetic position does not persist to low temperatures but suddenly disappears upon further cooling.
The suppression of the AFM ordering can be associated with the appearance of the intermediate phase and the reduction of the resistance, see Fig.~\ref{neutrons} and \ref{magnetism}.

 \section{Discussion}

The totality of diffraction studies in our and in previous \cite{Nakamura2013,Bertinshaw2019,Zhao2020,Cirillo2019} experiments remains puzzling and cannot be fully reconciled.
Most likely different experimental conditions - in particular concerning temperature control and measurement - have a
large impact on the real structure of the \cro \ crystals that carry finite current density below the metal-insulator transition. Our comprehensive diffraction
studies indicate that for the largest part of the studied current and temperature ranges the samples are structurally inhomogeneous presenting a complex
phase coexistence. Phase coexistence must even be expected in view of the $S$-shaped $j$-$E$ characteristics \cite{Volkov1969}.

Our experiments for currents perpendicular to the planes as well as all previous experiments agree concerning the
fact that only very high current densities applied at elevated temperature result in strong Bragg scattering that can be
associated with the high-temperature metallic state of \cro ,
i.e. a $c$ lattice constant close to 12.25\,\AA .
When ramping up a high current density of the order of \expj{10} and above, it seems unavoidable to heat the crystal and several groups report that
the $c$ value of the insulating phase increases close to $\sim$12.05\,\AA \cite{Nakamura2013,Bertinshaw2019,Zhao2020}, which is the largest $c$ value the insulating phase of \cro \ attains upon heating (without current) before it transforms to the metallic phase \cite{Friedt2001}.
Even though there is no doubt that metallic conduction is induced at temperatures well below the metal-insulator transition as can be seen in e.g. our time-resolved studies, any results with large current densities have to be considered with great care.
Our temperature dependent measurements at large current densities unambiguously show that even in this case the
metallic phase Bragg contributions rapidly diminish upon cooling. When analyzing the total scattering one must, however, keep the
spatial extension of the different phases in mind.
If the spatial extension of the metallic phase falls below the coherence length of the diffraction experiment, the metallic phase scattering will become broadened in $Q$ space. \add{For our XRD experiment with a conventional X-ray tube we can estimate the spatial coherence to be in the range of 40\,nm \cite{Bartzsch2017}, while the coherence length in the neutron experiments amounts to around 20\,nm .}
In our and all previous diffraction studies the use of tiny crystals \cite{Nakamura2013,Bertinshaw2019,Zhao2020} rendered it impossible to characterize such weak diffuse scattering.

The low-current state in \cro \ also exhibits a large negative magnetoresistance \cite{Sow2017} that is difficult to explain without magnetism. However, a similar observation in
an imperfect bulk-insulating topological insulator was attributed to the Zeeman effect on barely correlating current paths \cite{Breunig2017}.

Small metallic phase regions embedded in an insulating matrix will nevertheless impact the Bragg scattering of the insulating phase,
because the average $c$ lattice constant within the coherence volume increases.
Such effects can explain the behavior of the Bragg contributions of insulating and intermediate phases.
The $c$ constant of the insulating phase with finite current is very close to that of the current-free insulating phase,
and the small deviations can stem from some current-induced heating combined with the admixture of short-range (i.e. extension below the coherence length) metallic parts, but the diffraction studies cannot exclude an intrinsic effect due to e.g. homogeneous loss of orbital polarization.
In particular when ramping the current density up the inclusion of more and more metallic phase regions yields a continuous
increase of $c$. The crystal structure reported for a large current density of \Acm{10} at \K{130} \cite{Bertinshaw2019} exactly corresponds to that of
current-free \cro \ at \K{240} \cite{note-struc,Friedt2001}, therefore it can also be explained by metallic admixtures (combined with heating).
However, the DFT analysis proposing this structure as semimetallic \cite{Bertinshaw2019} can be questioned in view of the well established insulating properties of \cro \ at \K{240}.

In pure and in substituted crystals we find the intermediate phase under various conditions, and
evidence for a similar phase can also be found in the room temperature X-ray studies by Cirillo et al. \cite{Cirillo2019}.
The intermediate phase appears to be crucial for the understanding of the anomalous properties at low current densities,
which must result from some important intrinsic change. Our neutron experiments for
current densities of 1 to \Acm{2} suggest that the intermediate phase appears or becomes enhanced in the temperature range below 100\,K.
In addition we still find Bragg scattering at the position where AFM order contributes a Bragg peak in the normal insulating state.
This observation indicates that AFM ordering can coexist with current carrying parts in one crystal.
However, the AFM Bragg scattering disappears when the intermediate phase forms. Most likely the intermediate phase represents a more regular and more homogeneous arrangement compared to the modified insulating phase, but again a homogeneous loss of electronic order cannot be excluded
on the basis of the diffraction studies.
In the percolative picture of small metallic phase regions in an insulating matrix the intermediate phase is attributed to a stronger and more homogeneous content of metallic regions with most likely also a more regular arrangement.
The occurrence of the intermediate phase is accompanied with an enhanced conductance in agreement with a more regular arrangement.

The optical room-temperature studies by Zhang et al. \cite{Zhang2019} reveal a regular micro-stripe pattern of metallic phase and insulating phase regions.
Such a regular arrangement is common in martensitic phase transitions that create strong local strain \cite{Khachaturyan1983,Khachaturyan1991,Dec1993,Roytburd1993,Seo1998}.
In martensite transitions the domains may even form well-defined superstructures \cite{Khachaturyan1991,Kaufmann2010} arising from a regular arrangement of very small domain sizes causing superlattice reflections. A similar strain mechanism was also proposed to explain stripe-like phenomena in various transition-metal oxides \cite{Khomskii2001}.
Indeed the differences in the lattice constants between metallic phase and insulating phase in \cro \ without currents are huge. And these differences strongly
increase upon cooling. The low-temperature $c$ parameter of insulating \cro \ is $\sim$0.5\,\AA\,shorter than that of the metallic phase, and the difference in the orthorhombic $b$ parameter amounts to $\sim$0.25\,\AA. In contrast the $a$ parameter is almost identical.
The optical experiment examines an $a,b$ surface and finds stripes perpendicular to orthorhombic $b$, which is the
expected arrangement for martensitic domains that reduce the strong in-plane strain along $b$. Similar domain arrangements must, however, also exist perpendicular to the planes, as
the $c$ strain would be even larger. Both strain effects should result in a relative tilting of metallic phase and insulating phase domains. Since the relative strains will increase by a factor three upon cooling, the microstripe pattern will considerably change. The lattice strain seems to be the driving force for the rearrangement of the phase volumes and in particular for the occurrence of the intermediate phase at low temperatures.
We speculate that the intermediate phase is a more fine mesh of insulating and metallic parts, so that the average is better defined with intermediate lattice constants.
For this finer mesh there is enhanced mutual influence of one state on the other state, which explains the disappearance of AFM ordering.
However, studies with local probes are required to fully resolve the structural nature of the intermediate phase.
Considering the strain effects and the fact that the insulator-metal transition can be induced by only moderate hydrostatic pressure above 0.5\,GPa \cite{Nakamura2002,Steffens2005}, one may assume that \cro \ grown on a substrate causing tetragonal strain will even more easily undergo the insulator-metal transition at low temperature.

It also appears most important to understand the relation between the two mechanisms explained above: the formation of conducting filaments typically parallel to the current arising
from the general phase instability \cite{Volkov1969} and the impact of the crystal strain due to the large differences in lattice constants that will also favor domain formation
similar to the well known effects in martensite transitions \cite{Khachaturyan1983,Khachaturyan1991}. One  may only speculate how these two phenomena actually couple: under the influence of current, in a negative part of the $j$-$E$ curve of Fig.~2, the system becomes unstable. It first tends to form conducting filaments predominantly parallel to the current,
in which the insulator-metal transition and local heating imply strong structural changes. But then the strong strain imposes its own pattern in this inhomogeneous state leading to the formation of more complicated microstructures, with metallic inclusions not just parallel to the current, but largely determined by strain. Therefore, the current has to meander, which finally  leads to percolating network. The general picture of the origin and of the main characteristics of the current carrying state in  Ca$_2$RuO$_4$ are certainly related to both mechanisms.

\section{Conclusion}

X-ray and neutron diffraction studies as function of temperature and applied current density reveal
a complex real structure indicating that phase segregation and phase coexistence are essential for the physics
of the current carrying state in \cro . Evidence for phase segregation is deduced from the
S-shaped quasistatic $j$-$E$ curves and from two distinct time scales visible in pulsed transport measurements. 
When crystals are cooled with controlled current density we observe three
different contributions to the Bragg peaks that can be easily assigned to phases with distinct
$c$ lattice parameters. Except at very high current densities and at elevated temperatures, there are only minority phases corresponding to the high-temperature metallic state of \cro \ with $c$ lattice constants of the order of 12.25\,\AA.
The relative weight of these metallic contributions to the Bragg scattering continuously decreases upon cooling.
The suppression of the Bragg peaks associated with the long-$c$-axis metallic state, however, does not exclude the persistence
of such regions. The spatial extent of metallic regions can just become considerably lower than the coherence
length of the diffraction experiment \add{(of the order of 20 to 40\,nm)}.
For the lower current densities, below \expj{2}, for which anomalous low-temperature properties have been reported, metallic long-$c$ Bragg scattering is irrelevant.
The two other Bragg scattering contributions correspond to $c$ lattice parameters close to the much smaller values expected for the insulating state. One phase, labelled insulating phase, exhibits rather similar though significantly larger $c$ values compared to the current-free insulating state,
while an intermediate contribution, intermediate phase, appears with about $\sim$0.1\,\AA\,larger $c$ constants.
Upon cooling, heating and ramping the current densities up or down the phase ratio between insulating and intermediate states considerably varies, which is associated with changes in the resistance.
Most importantly we find evidence for AFM ordering even in samples carrying moderate current densities, but this AFM
ordering seems to be suppressed when parts of the sample transform to the intermediate state. 

The characters of the insulating and intermediate
phases in the current carrying states cannot yet be fully established, but it cannot be excluded that they simply arise from
phase coexistence and regular microarrangements of metallic and insulating regions.
The general tendency to phase segregate and the impact of the local strain are proposed to drive the
complex real structure in current carrying Ca$_2$RuO$_4$ at low temperature.

\begin{acknowledgments}
This work was funded by the Deutsche Forschungsgemeinschaft (DFG,
German Research Foundation) - Project number 277146847 - CRC 1238, projects A02, B02 and B04.
\end{acknowledgments}

\nocite{apsrev41Control}

\bibliographystyle{apsrev4-1}


%

\end{document}